\documentstyle[12pt]{article} 
\begin{document} 
\begin{center} 
COMMENT ON MAGNETIC MONOPOLE EXCITATIONS IN SPIN ICE 
\vskip 2cm 
Subir Ghosh\\ 
\vskip .3cm 
Physics and Applied Mathematics Unit,\\ 
Indian Statistical Institute,\\ 
203 B. T. Road, Calcutta 700108, India\\ 
\end{center} 
\vskip 2cm {\bf {Abstract:}} It has been proposed recently 
\cite{son} that excitations in Spin Ice can be of the form of 
magnetic monopoles that does not obey the Dirac Quantization 
Condition. It is also well known \cite{rj} that the above scenario 
leads to non-associativity among translation generators. It will 
be interesting to see how the monopole picture in Spin Ice 
survives in the light of the latter observation. 
\newpage 

Magnetic monopoles have become an enigma ever since Dirac 
\cite{dir} argued that the presence of only one of them can induce 
charge quantization through the product $eg$ where $e$ and $g$ are 
respectively the strengths of electric and Magnetic Monopoles 
(MM). Search of MM in High Energy Physics has not led to any 
positive outcome. In recent years, experimental observation of MM 
and its theoretical explanation \cite{fang} in Condensed Matter 
systems has created quite a stir. However the singular structure 
observed here is in momentum space. 

In a recent paper, Castelnovo, Moessner and Sondhi \cite{son} have 
proposed that some exotic properties of Spin Ice can be explained 
by postulating that the elementary excitations above the ground 
state are in the form of monopole-antimonopole pairs. In the approximation considered in \cite{son} the 
strings connecting the opposite poles comprise of dumbells of oppositely charged magnetic poles that replace each spin in the original system. These strings 
carry a small amount of energy and are in principle observable. Moreover, (and this is crucial to the present 
Comment), the Dirac Quantization Condition (DQC) is not imposed here and 
in fact the monopole charge can be varied continuously by 
applying pressure on the system \cite{son}. This can be contrasted to the 
conventional infinite Dirac string considered as a string of dipoles(see for example  \cite{jac}) 
which, being an artifact, has to be unobservable and this is tied up with gauge 
invariance of the quantum system of a charge in the presence of a 
monopole, which eventually leads to the DQC. 

A few years ago the problem of DQC was readdressed 
by Jackiw \cite{rj} in a lucid article (see also \cite{ber}). The beauty of the 
analysis \cite{rj} is that DQC is recovered 
in a {\it{gauge invariant}} way without taking recourse to a 
singular 
vector potential of Dirac string. On the other hand, the analysis 
is based on the simple fact that the quantum mechanics that we know 
today 
relies on the Hamiltonian formulation and lives in a Hilbert 
space. It is essential that the {\it{Jacobi identities}} involving the 
operators are satisfied and that the operators acting on the Hilbert space are 
{\it{associative}}. Jackiw's work \cite{rj} reveals that if one 
considers the semiclassical dynamics of a charge in an external MM 
field {\it{without}} imposing DQC, one is led to violation of Jacobi 
identities and non-associativity (among translation generators) 
in an essential way. From the very general nature of the arguments in 
\cite{rj}, it appears that they can be relevant in the monopole excitation model of Spin Ice mentioned above 
\cite{son}. 

Below we briefly reproduce the analysis of \cite{rj}. The gauge 
invariant dynamics of a massive ($m$) charge ($e$) in a 
non-dynamical magnetic field $B$ is given in terms of the particle 
position operator $\bf r(t)$: 
\begin{equation} 
\dot {\bf r}={\bf v } ~~;~~~ m\dot {\bf v}=\frac{e}{2c}({\bf v} 
\times {\bf B} -{\bf B} \times {\bf v}), \label{2} 
\end{equation} 
keeping in mind the non-commutative nature of $\bf v$ and $\bf 
B(r)$. The Hamiltonian, being independent of the magnetic field as 
it does not contribute to the particle energy, is 
\begin{equation} 
H=\frac{{\bf \pi} ^2}{2m}~~;~~{\bf \pi } \equiv m{\bf v}. 
\label{3} 
\end{equation} 
With the following non-trivial symplectic structure, 
\begin{equation} 
[r^i,r^j]=0~;~~[r^i,\pi ^j]=i\hbar \delta^{ij}~;~~[\pi ^i,\pi 
^j]=ie\frac{\hbar}{c} \epsilon^{ijk}B^k(\bf r), \label{4} 
\end{equation} 
one can reproduce the equations of motion (\ref{2}) in a 
Hamiltonian framework: 
\begin{equation} 
\dot {\bf r}=\frac{i}{\hbar}[H,{\bf r}]=\frac{\bf \pi}{m}~;~~\dot 
{\bf {\pi } }=\frac{i}{\hbar} [H,{\bf \pi }]=\frac{e}{2mc}({\bf 
\pi } \times {\bf B} -{\bf B} \times {\bf \pi } ). \label{5} 
\end{equation} 
However, Jacobi identity among the momenta yields, 
\begin{equation} 
\frac{1}{2}\epsilon^{ijk}[\pi ^i,[\pi ^j,\pi ^k]]=\frac{e\hbar 
^2}{c}\nabla .\bf B \label{6} 
\end{equation} 
which in fact vanishes for conventional magnetic fields 
$\bf\nabla . \bf B =0.$

It needs to be stressed \cite{rj} that the non-canonical (or non-commutative) phase space structure (\ref{4}) is forced on us if the Lorentz force law (\ref{2}) is to be derived consistently in a Hamiltonian framework. One can contrast this with the normal case $\bf\nabla . \bf B =0$ that leads to ${\bf B}=\nabla \times {\bf A}$. Now, with the identification $\pi ={\bf p} -\frac{e}{c}{\bf A}$ and the canonical phase space
$$[r^i,r^j]=0~;~~[r^i,p ^j]=i\hbar \delta^{ij}~;~~[p^i,p^j]=0,$$ it is trivial to derive non-commutative bracket 
$[\pi ^i,\pi 
^j]=ie\frac{\hbar}{c} \epsilon^{ijk}B^k(\bf r)$ and subsequently the Lorentz force law.

If one pursues further \cite{rj} one finds that for 
the translation generators, 
\begin{equation} 
T({\bf a})\equiv exp(-\frac{i}{\hbar}{\bf a}.{\bf \pi } 
)~~;~~~T^{-1}({\bf a})~{\bf r} ~T({\bf a})={\bf r} +{\bf a} 
\label{7} 
\end{equation} 
the Abelian composition law gets modified to 
\begin{equation} 
T({\bf a_1})T({\bf a_2}) = exp (-\frac{ie}{\hbar c}\Phi ({\bf 
r};{\bf a_1},{\bf a_2}))T({\bf a_1} +{\bf a_2})~;~~\Phi =({\bf 
a_1} \times {\bf a_2}).{\bf B} \label{8} 
\end{equation} 
This explicitly brings in to fore the loss of associativity with 
regard to translations, 
\begin{equation} 
(T({\bf a_1})T({\bf a_2}))T({\bf a_3})=exp(-\frac{ie}{\hbar 
c}\omega ({\bf r}; {\bf a_1},{\bf a_2},{\bf a_3}))T({\bf 
a_1})(T({\bf a_2})T({\bf a_3})), \label{9} 
\end{equation} 
with the offending factor being, 
\begin{equation} 
\omega = \int d{\bf S}.{\bf B} =\int dr ~\bf \nabla .{\bf B} 
\label{10} 
\end{equation} 
Clearly, $\omega $ measures the total magnetic flux coming out of 
the tetrahedron formed out of $\bf a_1,~\bf a_2,~\bf a_3$, with 
one vertex at $\bf r$. Now DQC can come to the rescue by imposing 
the condition, 
\begin{equation} 
\int dr {\bf \nabla } .{\bf B} =2\pi \frac{\hbar c}{e}N \label{11} 
\end{equation} 
since, for integer $N$,  the exponential factor reduces to identity. 
Furthermore the above also requires that the sources of magnetic 
field have point support, 
\begin{equation} 
{\bf B}=g\frac{{\bf r}}{r^3} ~~;~~{\bf \nabla } . {\bf B}=4\pi 
g\delta ^3({\bf r}). \label{12} 
\end{equation} 
Thus DQC $\frac{ge}{\hbar c}=\frac{N}{2}$ is restored. As for the 
Jacobi identity violation, it is not fully cured but is at least 
restricted only to the isolated points where the MMs are sitting \cite{rj}. 

Let us now return to the case at hand: Spin Ice. Loss of 
translation invariance will mean violation in the momentum 
conservation principle (as applied to a crystalline structure) and 
this can have direct experimental signatures. On the other hand, 
imposition of DQC will lead to a new quantization rule in the 
monopole magnitude of Spin Ice (see \cite{son}). Either way, 
further study of the new material Spin Ice, both in theoretical 
and experimental context, is bound to reveal new physics in a 
fundamental way.  

\end{document}